\documentclass{iaus}
\usepackage{graphicx}

\title[C/O Abundance Ratios and Dust Features in Galactic PNe] 
{C/O Abundance Ratios and Dust Features in Galactic Planetary Nebulae}

\author[Gloria Delgado-Inglada \& M\'onica Rodr\'iguez] 
{Gloria Delgado-Inglada \and M\'onica Rodr\'iguez}

\affiliation{Instituto Nacional de Astrof\'isica, \'Optica y Electr\'onica (INAOE), 
Apdo Postal 51 y 216, 72000 Puebla, Pue. Mexico, email: {\tt gloria@inaoep.mx,mrodri@inaoep.mx}}

\pubyear{2011}
\volume{283} 
\pagerange{}
\date{}
\jname{Planetary Nebulae: an Eye to the Future}
\editors{A. Manchado, L. Stanghellini \& D. Schoenberner, eds.}
\begin{document}

\maketitle

\begin{abstract}
The iron depletion factors found in Galactic planetary nebulae (PNe) span over
two orders of  magnitude, suggesting that there are differences in the grain formation
and destruction processes from object to object.  We explore here the relation
between the iron depletions, the infrared dust  features, and the C/O abundance
ratios in a sample of Galactic PNe.  We find that those objects with
$\mbox{C}/\mbox{O}<1$ show a trend of increasing depletions for higher values of
C/O, whereas PNe with $\mbox{C}/\mbox{O}>1$ break the trend and cover all the range
of depletions. Most of the PNe with $\mbox{C}/\mbox{O}<1$ show silicate features,
but several PNe with C-rich features have $\mbox{C}/\mbox{O}<1$, probably
reflecting the uncertainties associated with the derivation of C/O. 
PAHs are distributed over the entire range of iron depletions and C/O values. 

\end{abstract}

We have constrained the gaseous iron abundances in a sample of Galactic planetary nebulae
(PNe) with good quality optical spectra, using [Fe III] lines and the correction scheme 
of \cite{Rodriguez_05}. From these abundances we can estimate the depletion factors for Fe/O 
using $(\mbox{Fe}/\mbox{O})_\odot$ as a reference value for the total Fe/O abundance (in gas
and dust):
$[\mbox{Fe}/\mbox{O}] = \log(\mbox{Fe}/\mbox{O}) - \log(\mbox{Fe}/\mbox{O})_\odot$
(see \cite[Delgado-Inglada et al.\ 2009]{DelgadoInglada_09}; Delgado-Inglada 
\& Rodr\'iguez, 2011 in preparation). 
The range of iron depletions covers two orders of magnitude, suggesting 
differences in the grain formation and destruction processes from object to object.
These differences  could be related to the chemistry of the environment. 
The value of C/O in the atmospheres of asymptotic giant branch stars, 
the progenitors of PNe, determines whether C-rich or O-rich grains are formed. 
Therefore, the C/O abundance in the ionized gas of PNe is likely to be related 
to the different dust features observed in their infrared spectra. We study here 
the relation between the iron depletions, the infrared features, and the 
C/O ratios derived in a sample of Galactic PNe.

\begin{figure}[!th]
\begin{center}
\includegraphics[width=0.5\textwidth]{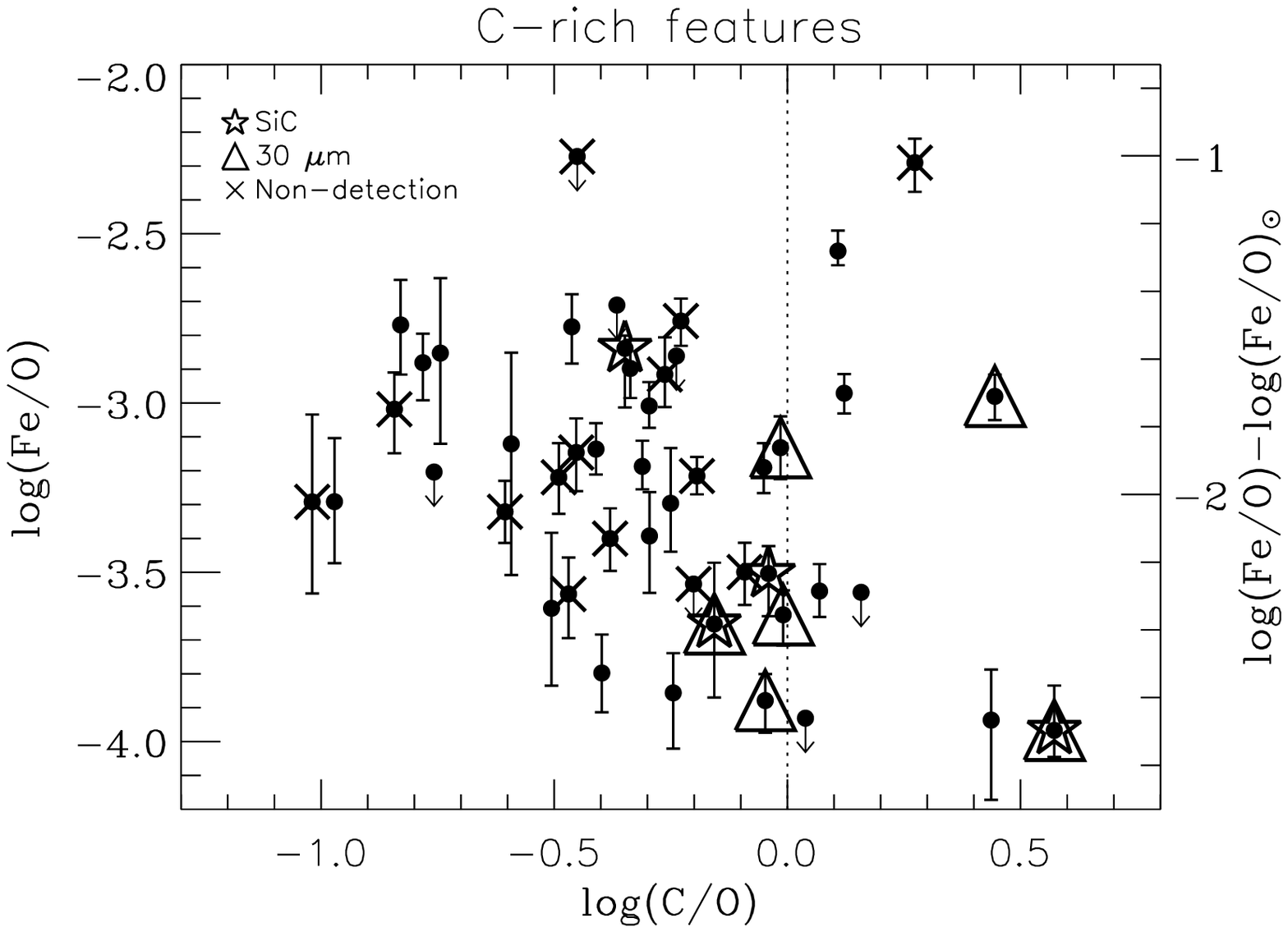} \hspace{-0.4cm}
\includegraphics[width=0.5\textwidth]{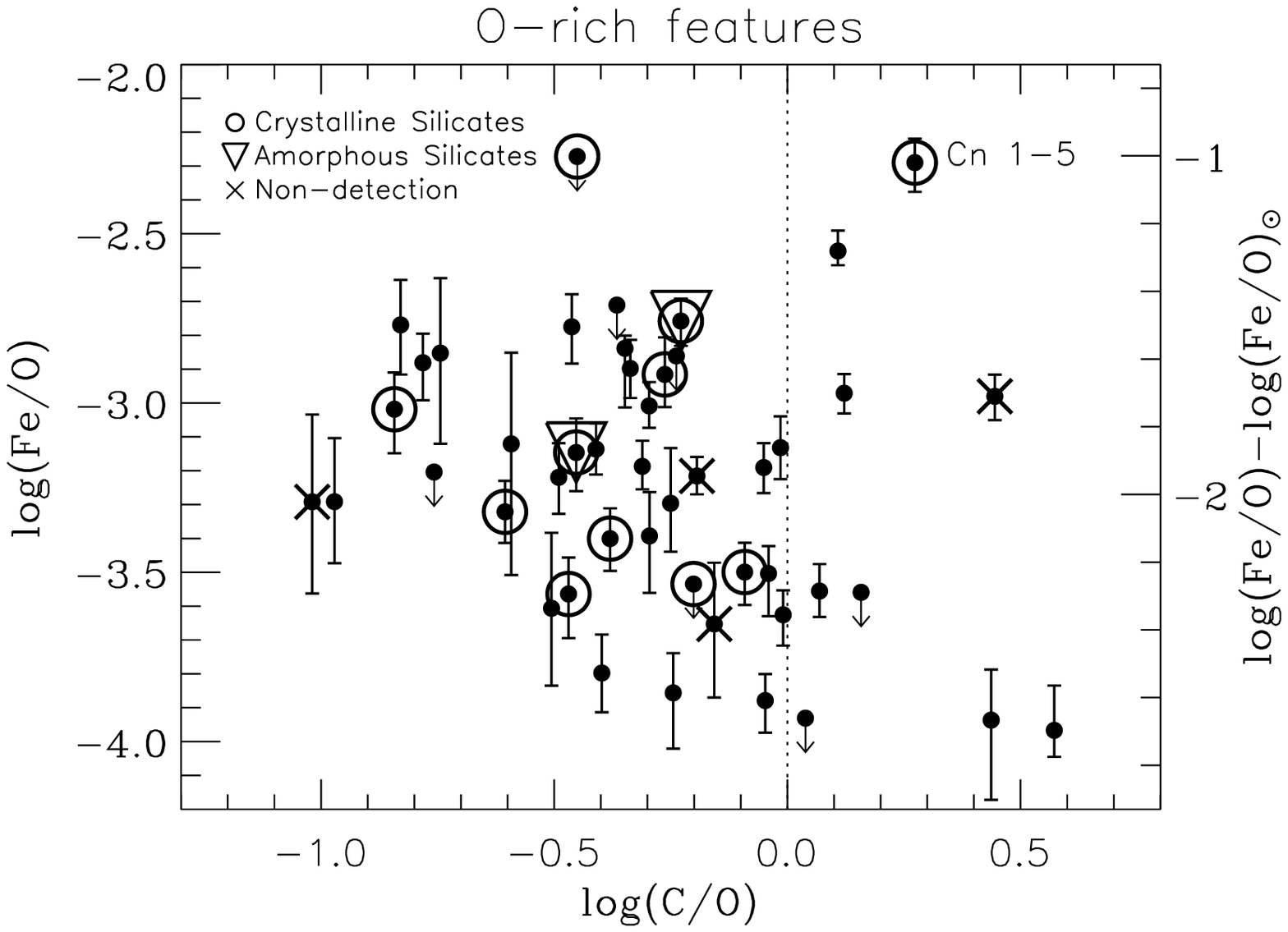} \\ 
\includegraphics[width=0.5\textwidth]{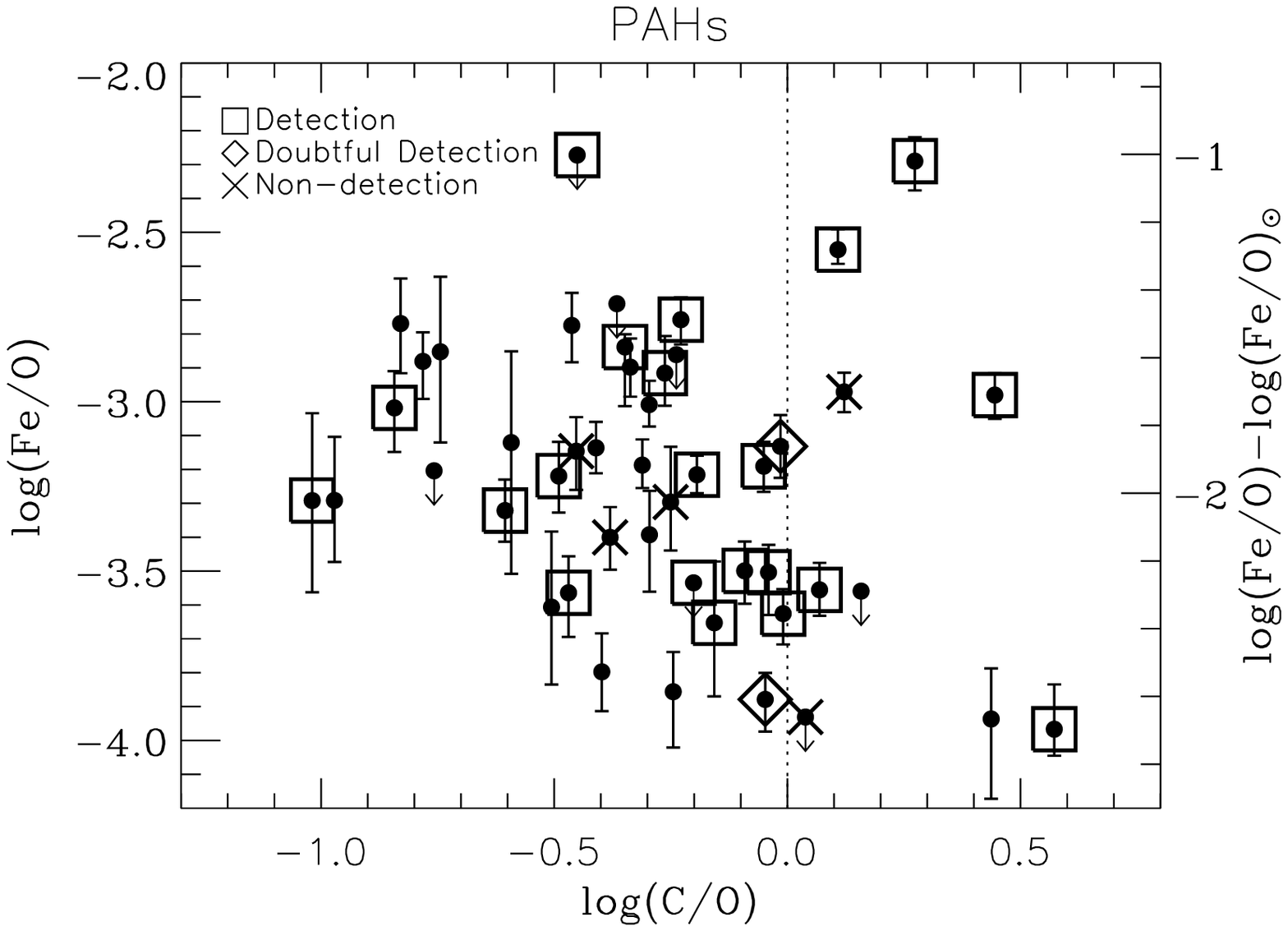}
\caption{Values of Fe/O (left axis) and the depletion factors [Fe/O] (right axis) as a 
function of C/O. We identify the dust features found in the sample PNe.}
\label{fig1}
\end{center}
\end{figure}

Figure \ref{fig1} shows the values of Fe/O and [Fe/O] as a function of the C/O
abundance ratios. The values of $\mbox{C}/\mbox{O}\simeq\mbox{C}^{++}/\mbox{O}^{++}$
were derived using  recombination lines for 48 PNe with available measurements. 
The use of these lines avoids the problems related to the combination 
of optical and ultraviolet observations, required when using collisionally 
excited lines. However, the C/O values are generally very uncertain, with 
discrepancies between the different methods (that use either recombination 
or collisionally excited lines, and different corrections for unobserved ions) 
of up to 0.6 dex. Figure \ref{fig1} also shows the different dust features found in
32 objects. We have reduced the available Spitzer/IRS spectra for our sample PNe,
and compiled other identifications from the literature (Delgado-Inglada \& Rodr\'iguez,
2011 in preparation).
The infrared features we use are the ones that are easily identified in the spectra
and that can be reliably associated with a C-rich or O-rich environment: PAHs, SiC,
the wide feature at 30 $\mu$m usually attributed to MgS
\cite[(Hony et al.\ 2002)]{Hony_02}, amorphous silicates and crystalline silicates. 

PNe showing the broad feature at 30$\mu$m and/or the SiC band are expected to
have $\mbox{C}/\mbox{O}>1$, but this is not the case for some of our objects,
suggesting that their C/O values are wrong. On the other hand, most, maybe all,
of the PNe with $\mbox{C}/\mbox{O}<1$ show silicate features.
We cannot rule out the presence of these features in two PNe shown as non-detections
in Figure \ref{fig1} due to the poor quality of their spectra. The other
non-detection, M1-20, shows carbonaceous dust features and is likely to have
$\mbox{C}/\mbox{O}>1$. On the other hand, Cn1-5 is
the only PN with $\mbox{C}/\mbox{O}>1$ that shows silicates. 
Either the C/O ratio is wrong for this nebula, or the silicates 
formed in a previous O-rich evolutionary stage of a now C-rich nebula. 
In such case, the silicates could be stored in a long-lived disk
\cite[(Waters et al. 1998)]{Waters_98}.
We find that PAHs are present in all the PNe with good quality infrared spectra 
(the halo PN DdDm~1 could be an exception). PAHs are distributed in the whole range of 
iron depletions and C/O values. The presence of PAHs is expected in 
C-rich environments; in O-rich PNe, PAH emission could arise in the surrounding 
interstellar medium or in a dense O-rich torus, where PAHs could form, 
as proposed by \cite[Guzm\'an-Ram\'irez et al.\ (2011)]{Guzman_11}.
The iron depletions increase with C/O for those objects with $\mbox{C}/\mbox{O}<1$,
whereas objects with $\mbox{C}/\mbox{O}>1$ cover the full range of iron depletions.
This could be due to gradual changes in the dust composition when C/O increases,
up to $\mbox{C}/\mbox{O}=1$, when there is an abrupt transition in dust composition.


\begin{thebibliography}{}

\bibitem[Delgado-Inglada \etal\ (2009)]{DelgadoInglada_09}
{Delgado-Inglada, G., Rodr\'iguez, M., Mampaso, A., \& Viironen, K.} 2009, \textit{ApJ}, 694, 1335 
\bibitem[Guzm\'an-Ram\'irez \etal\ (2011)]{Guzman_11}
{Guzm\'an-Ram\'irez, L. et al.} 2011, \textit{MNRAS}, 414, 1667
\bibitem[Hony, Waters, \& Tielens (2002)]{Hony_02}
{Hony, S., Waters, L. B. F. M., \& Tielens, A. G. G. M.} 2002, \textit{A\&A}, 390, 553
\bibitem[Rodr\'iguez \& Rubin (2005)]{Rodriguez_05}
{Rodr\'iguez, M., \& Rubin, R. H.} 2005, \textit{ApJ}, 626, 90
 \bibitem[Waters \etal\ (1998)]{Waters_98}
{Waters, L. B. F. M. et al.} 1998, \textit{A\&A}, 331, 61


\end{thebibliography}
\end{document}